\documentclass{cupconf}

  \checkfont{eurm10}
  \iffontfound
    \IfFileExists{upmath.sty}
      {\typeout{^^JFound AMS Euler Roman fonts on the system,
                   using the 'upmath' package.^^J}%
       \usepackage{upmath}}
      {\typeout{^^JFound AMS Euler Roman fonts on the system, but you
                   dont seem to have the}%
       \typeout{'upmath' package installed. cupconf.cls can take advantage
                 of these fonts,^^Jif you use 'upmath' package.^^J}%
      }
  \else
  \fi

  \checkfont{msam10}
  \iffontfound
    \IfFileExists{amssymb.sty}
      {\typeout{^^JFound AMS Symbol fonts on the system, using the
                'amssymb' package.^^J}%
       \usepackage{amssymb}%

      }{}
  \fi

  \IfFileExists{amsbsy.sty}
    {\typeout{^^JFound the 'amsbsy' package on the system, using it.^^J}%
     \usepackage{amsbsy}}
    {\providecommand\boldsymbol[1]{\mbox{\boldmath $##1$}}}

\newsavebox{\astrutbox}
\sbox{\astrutbox}{\rule[-5pt]{0pt}{20pt}}

\newcommand\etal{\mbox{\textit{et al.}}}

\newcommand\chandra{{\textit{Chandra}}}
\newcommand\einstein{{\textit{Einstein}}}
\newcommand\ergs{{$\rm{erg~s^{-1}}$}}
\newcommand\degr{$^{\circ}$}
\newcommand\mnras{\textit{MNRAS}}
\newcommand\apj{\textit{Ap. J.}}
\newcommand\apjs{\textit{Ap. J. Suppl}}
\newcommand\aap{\textit{A\&A}}
\newcommand\aaps{\textit{A\&A suppl.}}
\newcommand\pasj{\textit{PASJ}}

\usepackage{graphicx}

\title{First results from a \chandra~ survey of the 'Bar' region of the SMC}

\author[A. Zezas {\it et al.\/}]%
{A. Zezas$^1$%
J. C. McDowell$^1$, D. Hadzidimitriou$^2$, V. Kalogera$^3$,
G. Fabbiano$^1$, P. Taylor$^{1,4}$}

\affiliation{$^1$Harvard-Smithsonian Center for Astrophysics, 60
Garden Street, Cambridge, MA 02138, USA \\[\affilskip]
$^2$ Physics Dept. University of Crete, 71003, Herakleion, Greece \\[\affilskip]
$^3$ Northwestern University, 2145 Sheridan Rd, Evanston,  IL 60208\\[\affilskip]
$^4$ Dept. of Physics, Boston College, 140 Commonwealth Ave., Chestnut 
Hill, MA 02467
USA \\[\affilskip]}

\pubyear{2003}
\volume{}
\pagerange{}
\date{}
\begin{document}

\maketitle

\begin{abstract}
We present the first results from a \chandra~ survey of the central region of
the Small Magellanic Cloud.  We detect a total of 122 sources down to
a limiting luminosity of $\sim4.3\times10^{33}$\ergs (corrected for
Galactic $\rm{N_{H}}$), which is $\sim10$
times lower than in any previous survey of the SMC. We identify 20
candidate transient sources: eighteen previously known sources in this
area which are not detected in our observations, and two new bright
sources.   The spectral parameters
of the brightest sources indicate that they are X-ray binary pulsars.
The high spatial resolution of \chandra~ allows us to initially identify optical
counterparts for 35 sources, 13 of which are new identifications. 
\end{abstract}

\firstsection 
\section{Introduction}

 The Small Magellanic Cloud (SMC) is one of the prime objects to study 
the extragalactic X-ray binary populations because its
small distance and low Galactic extinction allows the detection of very
faint sources and the identification of their optical counterparts. For the same
reasons it is possible to determine its star-formation history much
more accurately than in more distant galaxies. 
This gives us the possibility to investigate the connection between
 star-formation and the X-ray binary populations.

 Studies of the stellar populations of the SMC show 
that its central region is dominated by a young stellar population from
a recent burst of star-formation 
which occurred between 50 and 10~Myr ago (e.g. Harris 2000; Maragoudaki 
\etal, 2001). Together with this population coexists a population of
older stars forming a uniform spheroidal distribution (e.g. Gardiner
\& Hadzidimitriou, 1992; Harris 2000). 

 The SMC has been observed with all major X-ray
observatories. \einstein~ detected over 70 sources down to a detection
limit of $\sim5\times10^{34}$ \ergs\footnote{Throughout this
paper the luminosities are in the 0.1-10.0~keV band, assuming an absorbed
($\rm{N_{H}=5.9\times10^{20}~cm^{-2}}$)  power-law model ($\Gamma=1.7$) 
  and are not
corrected for absorption, unless otherwise stated. The assumed
distance is 60~kpc (van den Bergh 2000).} over an area of 32\degr, 24 of
which have been identified as physically associated with the SMC (Wang
\& Wu, 1992).  ROSAT
performed two major surveys of the SMC, one with the PSPC (e.g. Haberl
\etal, 2000; Kahabka \& Pietsch, 1996)
and one with HRI (Sasaki \etal, 2000) detecting a total of 517 and
121 X-ray sources respectively. The detection limits of
these surveys vary across the observed area, with the flux of the
faintest sources being $\sim5\times10^{34}$~\ergs~ and
$\sim3\times10^{35}$~\ergs~ for the PSPC and the HRI 
surveys respectively. 

 The first hard X-ray survey of the SMC  (0.5-7.0~keV) was performed
with ASCA (Yokogawa \etal,  2000, 2003). This survey identified 106
individual sources, 5 of which were newly discovered pulsar
binaries (based on the detection of coherent pulsations), 
while 8 sources were classified as pulsar candidates
based on their hard spectra.
 Recently four fields in the  outer parts of the SMC were observed
with XMM-Newton  (Sasaki \etal, 2003) which identified two new pulsars
and two additional new X-ray sources.

\section{Description of the Survey}

 In order to study in detail the low-luminosity end of the  X-ray
binary population in an actively star-forming galaxy and its
connection to the stellar populations of the galaxy, we initiated a
\chandra~ survey of the ``Bar'' region of the SMC. This region hosted
the most recent starburst event in the SMC, between 50
and 10~Myr ago (Harris \etal, 2000). 

 We observed  5 individual fields, between
May and October 2002, with the \chandra~ ACIS-I camera
which provides a $16.9'\times16.9'$ field of view. Figure~1 shows the
observed fields overlaid on a DSS  image of the SMC.  The exposure times
 (7.6~ksec and  11.6~ksec) were chosen, based on 
an HI map of the SMC (Staveley-smith \etal, 1997), in order to achieve a
uniform detection limit of $\sim7\times10^{34}$\ergs (0.1-10.0~keV,
corrected for absorption) across the surveyed area, even for
sources located in the far edge of the SMC (i.e. are seen through
the maximum HI column density). 

\begin{figure}
  \includegraphics[width=9.5cm]{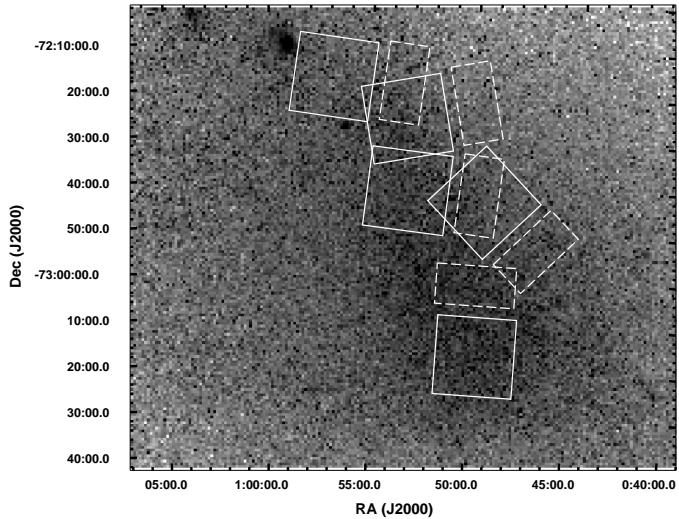}
  \caption{A DSS image of the central part of the SMC with the
outline of the 5 observed fields (solid lines show the ACIS-I array
which was at the aimpoint and dashed lines show the two ACIS-S CCDs.
 Each ACIS-I field is $16.9'\times16.9'$ wide.}\label{fig:wave}
\end{figure}

\section{First results}

 The data were analyzed following the standard procedures for the
analysis of \chandra~ data. Fig.~2 shows a full band (0.5-7.0~keV),
adaptively smoothed, exposure corrected, mosaic of the 5 observations. 
Images in soft
(0.5-2.0~keV), medium (2.0-4.0~keV) and hard (4.0-7.0~keV) bands were
searched for sources using the wavelets based {\textit{wavdetect}}
source detection algorithm. This search yielded between 21 and 32
sources per field at the $3\sigma$ level above the local background.
 The
 observed flux of the faintest sources is $\sim1.5\times10^{-14}$~$\rm{erg~s^{-1}~cm^{-2}}$
(0.1-10.0~keV) which corresponds to a luminosity of
\begin{figure}
\hspace{-2.0cm}
  \includegraphics[width=16.0cm]{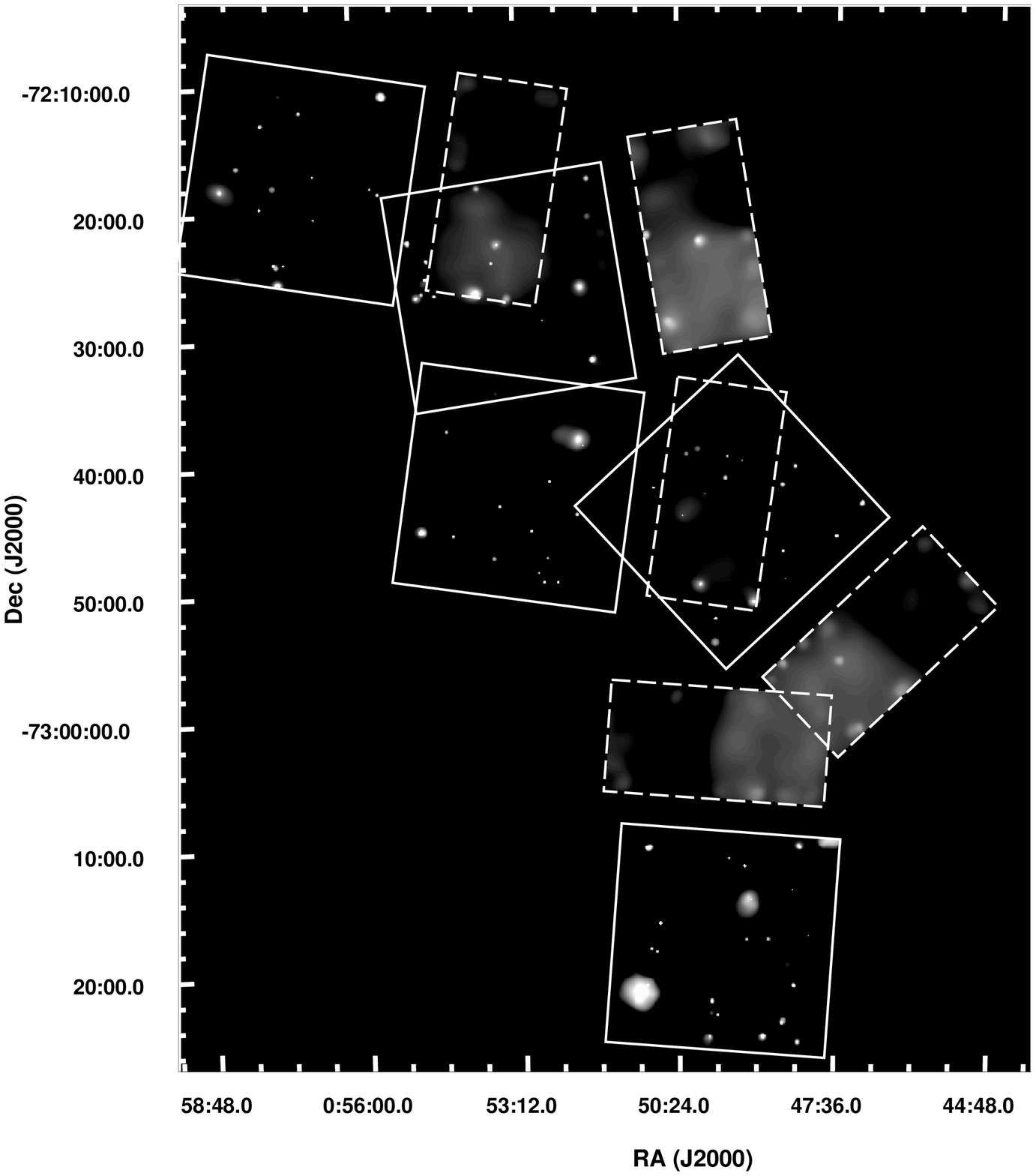}
  \caption{An adaptively smoothed, exposure corrected full band (0.5-7.0~keV) mosaic of
the \chandra~ observations. As in Fig.~1 solid lines show the ACIS-I array
 and dashed lines show the two ACIS-S CCDs. The enhanced diffuse
emission on one of the ACIS-S CCDs is due to the higher background of
the S3 CCD.  }\label{fig:contour}
\end{figure}
$\sim6.5\times10^{33}$~\ergs~ for the adopted distance of 60~kpc (van
den Bergh 2000).
This limit is $\sim10$ times lower than the limit of the ROSAT and the ASCA surveys.

 Based on the LogN-LogS distribution of the sources in the ChaMP
survey (Kim \etal, 2003) we expect to detect
 31 and 24 interlopers per field in the (0.5-2.0)~keV and
 (2.0-4.0)~keV  bands respectively. This is larger than the observed
number of sources by a factor of $\sim2$ and a factor of $\sim3$ for
the soft and the hard  band respectively (we detect 13-16 and 7-15 sources in
soft and medium bands respectively). 
 Even if all of the detected hard X-ray sources are not physically
associated with the SMC, this discrepancy indicates that the SMC
resides in a local minimum of the cosmic X-ray source distribution
(the fact that the excess of observed sources is visible in both the
medium and the soft band indicates that is not due to shadowing of the background
sources by the ISM of the SMC).  
 The association of at least 33\% of the detected X-ray sources with
stars in the SMC as well as their X-ray properties which are
consistent with pulsar X-ray binaries, suggest that a significant
fraction of them belongs to the SMC.

 The first results from the comparison of the \chandra~ data with the
ROSAT and ASCA surveys indicate that at least two of the detected sources are
transient sources.  Moreover, 18 sources previously detected by ROSAT
or ASCA were not detected in our surveys despite the 10-fold higher
sensitivity, indicating that they may also be X-ray
transients.  

Based on their extent and soft colour we identify at
least 5 candidate supernova remnants, 3 of which were previously known. 
In one of them (previously detected by ROSAT and \einstein)  we find a 
central, weak ($\rm{L_{X}=1.6\times10^{34}}$~\ergs, assuming isotropic
emission), source with a hard spectrum 
possibly associated with a pulsar in the center of this SNR.

All 14 point-like sources for which it is possible to perform
spectral fitting have hard X-ray spectra ($\Gamma<2.0$) indicative of
pulsar X-ray binaries.  Five of them were not detected in the ASCA
survey and two were not detected in the ROSAT surveys either. 
For the sources with previously published ASCA spectra we find
consistent results within the errors. However, the fluxes we derive
are systematically lower than the ASCA fluxes; in some cases this
might be due to
the fact that the  ASCA sources are resolved in several
\chandra~ sources.   

 Comparison  with the MOA catalogue of eclipsing binaries (Bayne
\etal, 2002) did not yield any coincidences. On the other hand
comparison with the optical source catalogue from the photometric survey of
Massey \etal~ (2002) gave 43 associations with 37 \chandra~ sources
 (the chance coincidence rate is 50\%). One
of these sources is of B2 spectral type while the other one is
classified as B-extreme type star (Massey 2002).  For thirteen
previously detected X-ray sources we identify   for the first time  
 candidate optical counterparts.

 In a nutshell the \chandra~ survey of the central part of the SMC
provides a view of the X-ray source populations in the youngest
regions of this galaxy with a ten-fold improvement in sensitivity over 
the previous X-ray surveys. The first results show that the majority
of the bright sources (most of which were previously known) are High
Mass X-ray binaries.  Key for the association of the detected
sources with the SMC is the identification of their optical
counterparts. For 1/3 of the sources we have identified optical
counterparts (13 of which are new) while with future optical follow-up
observations will pursue the identification of counterparts for the remaining
 sources, which will allow us to determine their nature and construct
the most complete (in terms of luminosity limit) sample of X-ray
binary in an external galaxy so far.

\begin{acknowledgments}
 We acknoweledge support by NASA contract NAS8-39073 (CXC),  NASA
Grants GO2-3127X and NAG5-13056 and  the NSF through the
Research Experiences for Undergraduates (AST-9731923) program.
\end{acknowledgments}



\begin{thebibliography}{}

\bibitem[van den Bergh(2000)]{2000glg..conf.....V} van den Bergh, S.\ 2000, 
The galaxies of the Local Group, by Sidney Van den Bergh.~Published by 
Cambridge, UK: Cambridge University Press, 2000 Cambridge Astrophysics 
Series Series, vol no: 35, ISBN: 0521651816.,  

\bibitem[Gardiner \& Hatzidimitriou(1992)]{1992MNRAS.257..195G} Gardiner, 
L.~T.~\& Hatzidimitriou, D.\ 1992, \mnras, 257, 195 

\bibitem[Haberl, Filipovi{\' c}, Pietsch, \& 
Kahabka(2000)]{2000A&AS..142...41H} Haberl, F., Filipovi{\' c}, M.~D., 
Pietsch, W., \& Kahabka, P.\ 2000, \aaps, 142, 41 

\bibitem[Harris(2000)]{2000A&AS..142...41H} Harris J., R., 2000,
Ph. D. Thesis, University of California, Santa Cruz 

\bibitem[Kahabka, Pietsch, Filipovi{\' c} , \& 
Haberl(1999)]{1999A&AS..136...81K} Kahabka, P., Pietsch, W., Filipovi{\' c} 
, M.~D., \& Haberl, F.\ 1999, \aaps, 136, 81 

\bibitem[Kahabka \& Pietsch(1996)]{1996A&A...312..919K} Kahabka, P.~\& 
Pietsch, W.\ 1996, \aap, 312, 919 

\bibitem[Kim]{} Kim, D. W. {\textit{et. al.}}, 2003,  \apj, in press (astro-ph/0308493)

\bibitem[Maragoudaki et al.(2001)]{2001A&A...379..864M} Maragoudaki, F., 
Kontizas, M., Morgan, D.~H., Kontizas, E., Dapergolas, A., \& Livanou, E.\ 
2001, \aap, 379, 864 

\bibitem[Massey(2002)]{2002ApJS..141...81M} Massey, P.\ 2002, \apjs, 141, 
81 

\bibitem[Sasaki, Haberl, \& Pietsch(2000)]{2000A&AS..147...75S} Sasaki, M., 
Haberl, F., \& Pietsch, W.\ 2000, \aaps, 147, 75 

\bibitem[Sasaki, Haberl, \& Pietsch(2002)]{2002A&A...392..103S} Sasaki, M., 
Haberl, F., \& Pietsch, W.\ 2002, \aap, 392, 103 

\bibitem[Sasaki, Pietsch, \& Haberl(2003)]{2003A&A...403..901S} Sasaki, M., 
Pietsch, W., \& Haberl, F.\ 2003, \aap, 403, 901 

\bibitem[Staveley-Smith et al.(1997)]{1997MNRAS.289..225S} Staveley-Smith, 
L., Sault, R.~J., Hatzidimitriou, D., Kesteven, M.~J., \& McConnell, D.\ 
1997, \mnras, 289, 225 

\bibitem[Wang \& Wu(1992)]{1992ApJS...78..391W} Wang, Q.~\& Wu, X.\ 1992, 
\apjs, 78, 391 

\bibitem[Yokogawa et al.(2000)]{2000ApJS..128..491Y} Yokogawa, J., 
Imanishi, K., Tsujimoto, M., Nishiuchi, M., Koyama, K., Nagase, F., \& 
Corbet, R.~H.~D.\ 2000, \apjs, 128, 491 

\bibitem[Yokogawa et al.(2003)]{2003PASJ...55..161Y} Yokogawa, J., 
Imanishi, K., Tsujimoto, M., Koyama, K., \& Nishiuchi, M.\ 2003, \pasj, 55, 
161 

\end{thebibliography}
\end{document}